\documentclass[iop,preprint2]{emulateapj}  
\usepackage{graphicx,amsmath,url}
\usepackage{times}
\usepackage{natbib}

\usepackage{bm}

\begin{document}

\title{Mineral Processing by Short Circuits in Protoplanetary Disks}
\shorttitle{Mineral Processing by Short Circuits}
\author{Colin~P.~McNally\altaffilmark{1,2,3}, Alexander Hubbard\altaffilmark{2}, Mordecai-Mark Mac Low\altaffilmark{2,3}, Denton~S.~Ebel\altaffilmark{4}, \& Paola D'Alessio\altaffilmark{5}}
\shortauthors{ C.~P.~McNally et al.}
\altaffiltext{1}{Niels Bohr International Academy, Niels Bohr Institute, Copenhagen, Denmark}
\altaffiltext{2}{Department of Astrophysics, American Museum of Natural History, New York, NY 10024-5192, USA}
\altaffiltext{3}{Department of Astronomy, Columbia University, New York, NY 10027, USA}
\altaffiltext{4}{Department of Earth and Planetary Sciences, American Museum of Natural History, New York, NY 10024-5192, USA}
\altaffiltext{5}{Centro de Radioastronom\'{\i}a y Astrof\'{\i}sica,
  Universidad Nacional Aut\'onoma de M\'exico, 58089 Morelia, MICH, M\'exico}
\email{cmcnally@nbi.dk, ahubbard@amnh.org, mordecai@amnh.org,
  debel@amnh.org, p.dalessio@crya.unam.mx}

\begin{abstract}
Meteoritic chondrules were formed in the early Solar System by
brief heating of silicate dust to melting temperatures.
Some highly refractory  grains (Type B calcium-aluminum rich inclusions, CAIs)
also show signs of transient heating. 
A similar process may occur in other protoplanetary disks,
as evidenced by observations of spectra characteristic of crystalline silicates.
One possible environment for this process is the turbulent magnetohydrodynamic flow
thought to drive accretion in these disks.  Such flows generally
form thin current sheets,
which are sites of
magnetic reconnection and dissipate the magnetic fields amplified by a disk dynamo.
We suggest that it is possible to heat 
precursor grains for chondrules and other high-temperature minerals
in current sheets that have been concentrated by our 
 recently described  short-circuit instability.
We extend our work on this process by including 
 the effects of radiative cooling, taking into account the temperature
dependence of the opacity;
and by examining current sheet geometry in three-dimensional, global models of
magnetorotational instability.
We find that temperatures above 1600~K can
be reached for favorable parameters that match
the ideal global models. 
This mechanism could provide an efficient means of tapping the gravitational
potential energy of the protoplanetary disk to  heat grains strongly enough to form high-temperature minerals. 
The volume-filling nature of turbulent magnetic reconnection
is compatible with constraints from chondrule-matrix
complementarity, chondrule-chondrule complementarity, 
the occurrence of igneous rims, and compound chondrules.
The same short-circuit mechanism may 
perform other high-temperature mineral processing in protoplanetary disks
such as the production of crystalline silicates and CAIs.
\end{abstract}


\section{Introduction}
\label{sec_intro}
The meteoritical record indicates that chemical and mineralogical evolution of protoplanetary disks
depends on local heating processes.
The ubiquity of chondrules in meteorites, along with the relation between chondrules and matrix (including chondrule-matrix elemental complementarity),
appears to demand a heating mechanism that acts 
locally, near the assembly point of chondritic meteorites.
Type B calcium-aluminum rich inclusions (CAIs), found in CV chondrites, also appear to have experienced transient heating, though under different conditions.

The basic challenge in producing chondrules is heating precursor grains
with   radii 
$\sim1~\mathrm{mm}$ from below $1000~\mathrm{K}$ to
$\sim 1800~\mathrm{K}$, and then cooling them at rates of 100--1000~K~hour$^{-1}$,
significantly slower than expected 
for free radiation to a cold background
\citep{1990Metic..25..309H,1990GeCoA..54.3537L,1998GeCoA..62.2725C,2005ASPC..341...15S,2006mess.book..253E}.
It further appears that heating to
$\sim 2000~\mathrm{K}$,  followed by cooling to  
$1500-1800~\mathrm{K}$  at  $\sim 5000$~K~hour$^{-1}$
is necessary to form barred textures \citep{2006mess.book..383C}.
Particularly low starting temperatures around 650~K may be required to retain the sodium and
other volatile elements that are found in some chondrules
\citep{2008Sci...320.1617A}, as these volatile elements will be
released when the chondrules are melted. 
Regardless of the starting temperature, 
strong clustering of dust grains and rapid processing is required to avoid
depletion  \citep{2008Sci...320.1617A}.

Examination of the chondrules and matrix in several chondritic meteorites
has shown that the two constituents  appear
complementary in that, though the elemental makeup of each varies
substantially, the bulk chemical compositions 
of chondrites made up of both always reflects the elemental ratios of the sun
\citep{1984GeCoA..48.1663K,2005PNAS..10213755B,2010E&PSL.294...85H}.
Indeed, chondrules with high variability in Fe/Si may themselves be
complementary to each other, combining into parent asteroids with solar Fe/Si ratios
\citep{2008M&PS...43.1725E}.
This suggests that chondrules must have formed from local heating of matrix material.
Grains from other stars (presolar grains) and insoluble organic matter
found in the matrices of most primitive chondrites could not have survived the heating
experienced by chondrules
\citep{2002GeCoA..66..661M,2006mess.book..383C},
establishing that the heating events were intermittent and did not
process all available material.
Many chondrules experienced multiple heating events, 
as shown by the occurrence  of igneous rims
\citep{2012M&PS...47.1176J},
and collisions with partially molten and solid
droplets to produce compound chondrules \citep{2004M&PS...39..531C}.
This supports the hypothesis that the chondritic meteorites are formed soon after
chondrule melting, out of a dense cluster of dust grains only a fraction of
which are melted.  Such a scenario requires a highly localized heating source.

Type B CAIs are remelted highly refractory inclusions.
They show evaporation of Mg and Si due to melting and evaporation in
reheating events after condensation 
\citep{2000GeCoA..64.2879G,2002GeCoA..66..521R,2007E&PSL.257..497S}.
Recent evidence suggests that the epochs of CAI and chondrule formation overlap 
\citep{2007ApJ...671L.181M,2009LPI....40.2006Y,Connelly2012}.

A possible energy source for this localized heating is the
      magnetorotational instability (MRI),
which 
      taps the orbital energy in the disk's differential rotation to
      drive magnetized turbulence, transporting angular momentum
      outward and allowing accretion to occur
\citep{Velikhov59,1960PNAS...46..253C,1991ApJ...376..214B}
The MRI
    driven turbulence dissipates energy as heat.
However, the heating occurs intermittently, not uniformly.  MHD
turbulence quite generally forms current sheets
\citep{1972ApJ...174..499P,1994ISAA....1.....P,1997PhR...283..227C}, that dissipate
energy faster than the average rate. Magnetic reconnection can
occur in such current sheets, enhancing their local heating rate in
small regions throughout the current sheets \citep[For a review,
see][]{2010RvMP...82..603Y}.  
\citet[][hereafter Paper~I]{2012ApJ...761...58H} showed that if the ionization depends strongly on
the temperature, the dropping resistivity within the region heated
by a current sheet can drive an instability effectively resulting
in a short circuit, leading to strong localized heating.

One of the main diffusive processes
limiting where MRI occurs in a
protoplanetary disk is Ohmic resistivity, which is in turn a function of the
degree of ionization of the gas.
Over much of the disk, estimates suggest that the ionization near the
midplane is too low for the
flow to be unstable to the MRI.  It is hence
expected that magnetically inactive dead zones exist in these dense,
cool, and almost entirely neutral regions
\citep{1996ApJ...457..355G}.  
The secular movement of these dead zones through the disk may
 temporarily suppress the MRI, and thus transient heating,
 potentially explaining the broad range of observed chondrule ages.

The most common treatment of ionization of a protoplanetary disk includes two
main components, thermal ionization of alkali metals (most importantly
potassium), and nonthermal ionization from radiation (stellar 
ultraviolet light and 
x-rays, cosmic
rays, and radionuclides; \citealt{2011ARA&A..49..195A}).
It has also been proposed that non-ideal MHD effects beyond Ohmic
resistivity limit MRI in various regions of the disk \citep{1994ApJ...421..163B,2011ApJ...739...50B}, and that
the smallest dust grains can themselves be the dominant charge carriers in the
outer disk \citep{2011ApJ...739...51B}.

This letter first presents an analysis of typical current sheet geometries in
MRI driven turbulence, and then presents one-dimensional models
including radiative transfer, that demonstrate that
the short-circuit instability first described in Paper~I
can produce sufficiently 
high temperatures in these current sheets
 to implicate them in the formation of high-temperature minerals.
This instability is controlled by the rapid drop in
resistivity with increasing temperature that results from thermal
ionization of potassium  at $T > 1000~\mathrm{K}$.

Other mechanisms for melting chondrule precursors by means of magnetic reconnection
have been proposed.
\citet{1979GeoRL...6..677S} suggested that beams of relativistic
electrons emerging from reconnection regions
above or below the dense disk
could heat chondrule precursors. 
The reconnecting magnetic fields were hypothesized to originate either from the Sun,
or a disk dynamo.
Along  with a model for the intermittency of reconnection in a
protoplanetary disk, \citet{2010MNRAS.404.1903K} proposed that magnetic
reconnection regions
could drive
shocks strong enough to melt chondrule precursors.
Formation of chondrules by ambipolar diffusion dominated reconnection was
explored by \cite{2004ApJ...606..532J}.
However, that mechanism is strongly suppressed by even a small guide field in
the reconnection region \citep{2003ApJ...590..291H}.

The electrical discharge mechanism of \citet{2005ApJ...628L.155I} and \citet{2012ApJ...760...56M} 
may interact with the short-circuit instability, being triggered when the current sheet is sufficiently intensified.

In Section~\ref{sec_global} we describe the bounding geometries of current
sheets analyzed in a global zoom-in simulation of MRI driven turbulence.
In Section~\ref{sec_onedim} we demonstrate that the short-circuit instability
 can heat these current sheets to
chondrule-forming temperatures, and that the temperature variation of 
opacity can lead to a self-regulated form of the instability.
We discuss the results, implications, and what further work must be done to
establish short circuits  as a heating mechanism in protoplanetary disks in
Section~\ref{sec_discussion}.

\section{Current Sheet Geometries} \label{sec_global}

\begin{figure}
\plotone{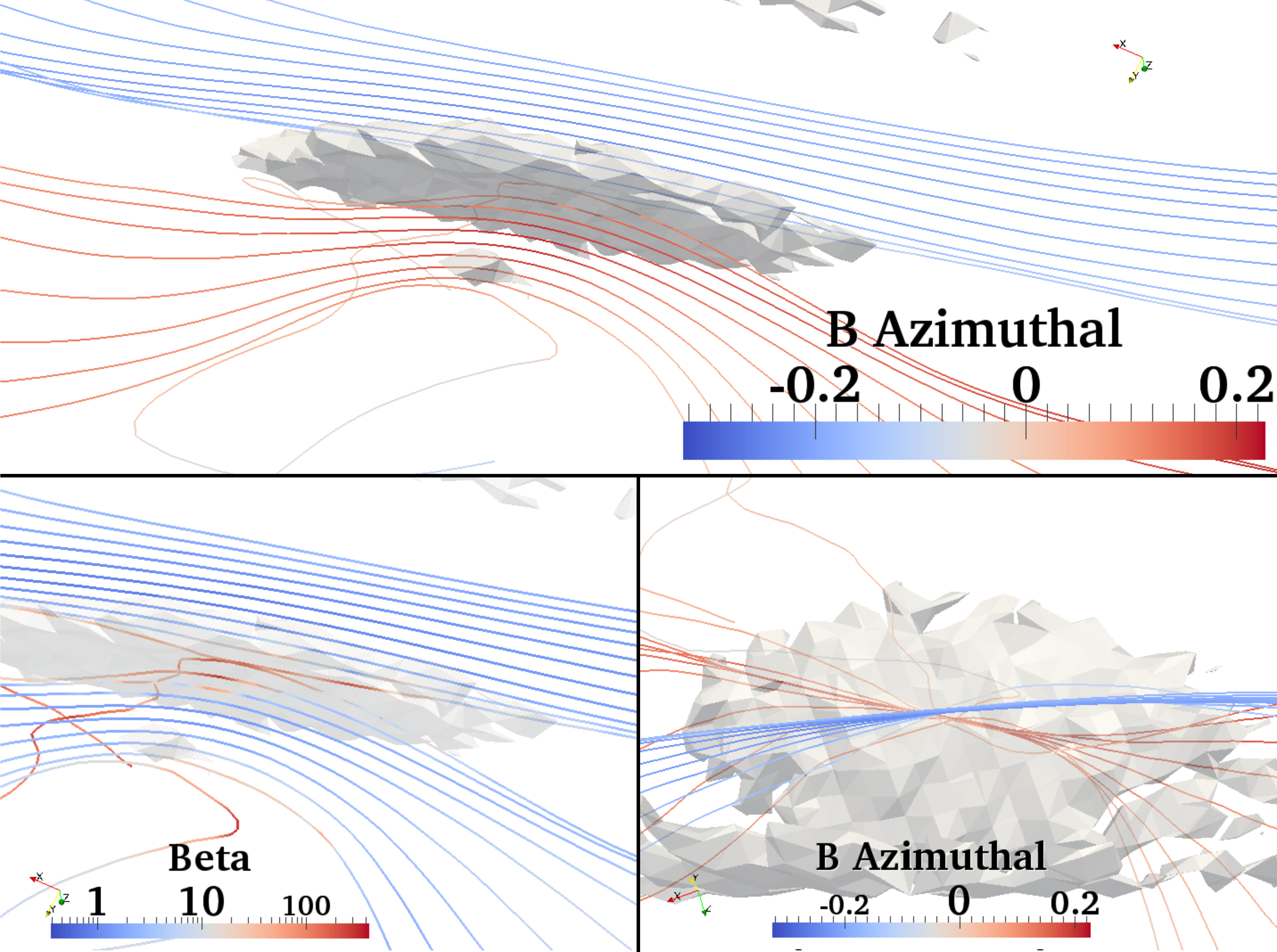}
\caption{A typical reconnection region found in the global MRI model.
The gray isosurface outlines high current density.
{\em Upper:} Streamlines show magnetic field, colored by azimuthal component.
The current sheet is bordered by oppositely directed azimuthal flux tubes. 
{\em Lower Left:} The same viewpoint as the upper panel, but the streamlines
are now colored by plasma $\beta$. The value of $\beta$ in the flux tubes bordering the
current sheet is $\sim 1$, and at least $\sim 500$ in the current sheet. 
{\em Lower Right:} Same quantities as the upper panel, with the viewpoint
rotated to view down the short axis of the current sheet.  The directions of
the magnetic field in the bounding flux tubes can be seen to be nearly
completely opposed.
Field lines originating from points inside the current sheet where the field is weak wander significantly away from the directions of the bounding flux tubes.
\label{reconn}}
\end{figure}

\begin{figure*}
\plotone{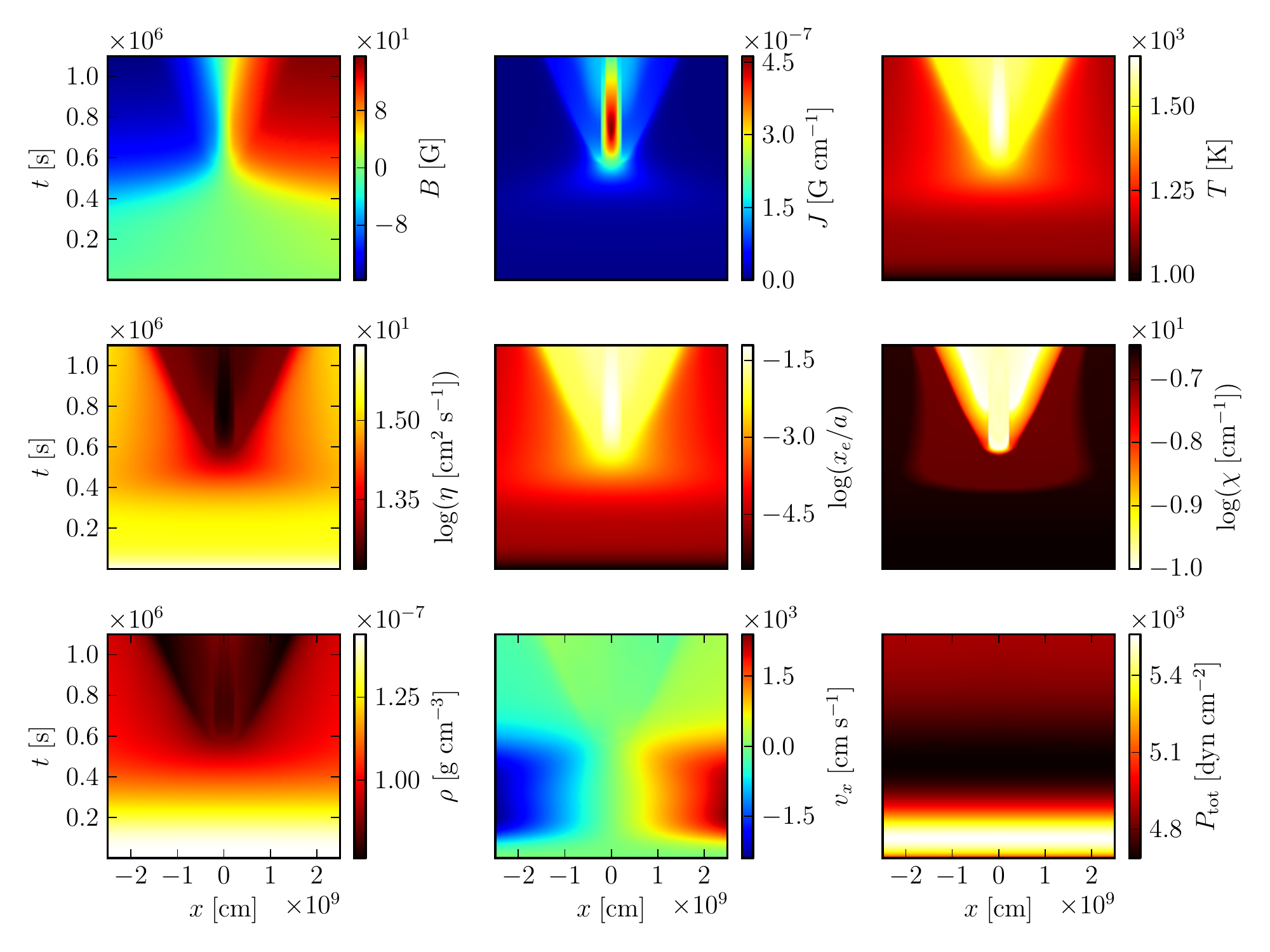}
\caption{Heating of a current sheet bounded by $\beta=1.5$, starting at $T_0=850$~K. 
The current sheet narrows though the short-circuit instability as the resistivity $\eta$ drops.
The peak temperature reached is $T \sim 1650\mathrm{\ K}$.
Shown are magnetic field $B$, current density $J$,  temperature $T$, resistivity $\eta$, potassium ionization fraction $x_e/a$,  inverse mean photon free path $\chi$, density $\rho$, velocity $v_x$, 
and total pressure $P_\mathrm{tot}$, with respect to spatial position $x$ and time $t$.
The results are plotted  mirrored about $x=0$ following the boundary conditions, and only the central section of the spatial domain is shown.
}
\label{fig_class2r3}
\end{figure*}

\begin{figure*}
\plotone{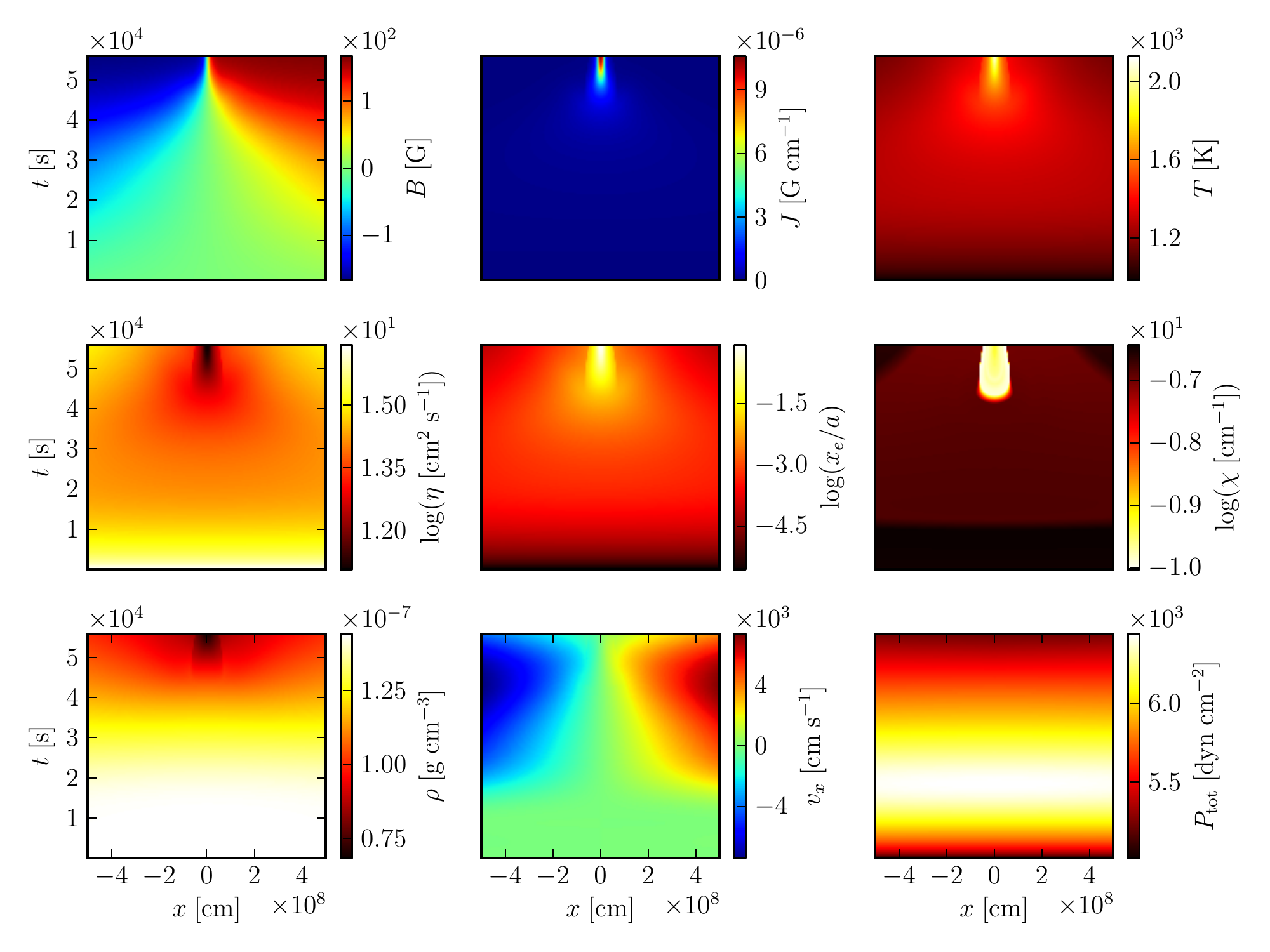}
\caption{
A thinner current sheet
  with width $10^{10}~\mathrm{cm}$, with all other parameters identical to
   Figure~\ref{fig_class2r3}, in which the potassium ionization saturates.
   The results are plotted  mirrored about $x=0$, and only the central section of the spatial domain is shown.
}
\label{fig_class2r4}
\end{figure*}

 We used Phurbas, an adaptive, Lagrangian, meshless, MHD code
\citep{2012ApJS..200....6M,2012ApJS..200....7M}, to perform
a global, unstratified simulation of the MRI in a protoplanetary
disk, in order to examine current sheet strength and geometry.
An MRI-unstable cylindrical Keplerian shear flow with
maximum radius $2.6~\mathrm{AU}$ is placed in the center of a vertically periodic
computational volume $40\times 40\times0.4~\mathrm{AU}$.
An isothermal equation of state is used, with sound speed corresponding to a
Keplerian disk with  aspect ratio of $0.1$, typical of protoplanetary disks. 
Initially, a  vertical magnetic field is
imposed, with a strength such that the critical MRI wavelength is one
quarter of the vertical height of the volume. 

Once the MRI-driven flow is reasonably saturated (as indicated by the evolution
of the magnetic tilt angle),
      two nested,
crescent-shaped,
   more highly refined
regions that
rotate with the flow are added, centered at a radius $1~\mathrm{AU}$,
       and extending $\sim\pi/2$ radians.  These
achieve resolution $1.7$ and $3.4$ times greater than the
base flow.
      The Phurbas resolution parameter $\lambda = 10^{-2}$~{AU}
      in the lowest resolution region, and decreases
to $2.8\times10^{-3}~${AU} 
         in the highest resolution region.
This is the region where current sheets were examined.
Further details are provided in Chapter 6 of \citet{mcnally12}.

In Figure~\ref{reconn} we show a typical high current-density region, which takes
the form of an approximately two-dimensional current sheet sandwiched by large,
nearly azimuthal flux tubes, whose magnetic field approaches the maximum field
at their orbital position.  
The bottom panels show that there is very little perpendicular guide
field passing through the center of
the sheet, as the magnetic fields on the opposite sides are nearly perfectly
opposed.
In subsonic MHD turbulence, the expectation is that the peak
currents occur at the tail of a turbulent cascade: spatially intermittent,
short in duration and length scale, and randomly oriented.  Peak magnetic
fields generated by the MRI are, however, near equipartition, which may explain
the large scale high current density regions.

We find that the current sheets retain the same general structure as 
the resolution is increased, while becoming thinner and more concentrated.
The total, gas plus magnetic, pressure is approximately constant across these
sheets, so in this locally isothermal simulation the sheets are regions of high
density.
From these results we conclude that a Harris-like current sheet geometry
\citep{Harris1962} as invoked in Paper~I, bounded by 
magnetic fields with magnetic pressure comparable to thermal
pressure, 
is well motivated for the study of Ohmic heating in protoplanetary disks.
This configuration uses a $\tanh$ profile for the magnetic field across the
current sheet, and the total pressure is set constant by satisfying an
adiabatic-hydrostatic condition with the temperature and density.

\section{One Dimensional Models} \label{sec_onedim}
We extended the implicit, one-dimensional, MHD code used in Paper~I, 
which uses sixth-order finite differences
on a logarithmic grid, and the CVODE package \citep{Hindmarsh:2005:SSN:1089014.1089020}
for time integration.  The extensions
include radiative transfer, and a full Saha-equation based treatment
of the ionization of potassium (with abundance relative to hydrogen of $a=10^{-7}$) and background nonthermal ionization.
The nonthermal ionization is parameterized following equation~(14) of
\citet{2002MNRAS.329...18F}, with the ionization rate 
$\zeta=10^{-16}~\mathrm{s^{-1}}$,  an order of magnitude higher than the canonical  value from
 cosmic ray ionization $\zeta=10^{-17}~\mathrm{s^{-1}}$. 
This nonthermal ionization component is used to maintain low enough resistivity to hold
our artificial current sheets
in place, in the absence of an active turbulent flow. This slightly increases
the onset temperature of the instability and slightly 
weakens its growth rate, so that our model establishes a lower limit
to susceptibility to the instability.

The code used in this Letter includes
a radiative heating and cooling term calculated with the ray tracing scheme presented in
\citet{2006A&A...448..731H}.  This radiative cooling is
along one dimension, as the global simulations have shown that the transverse dimensions
of the current sheets are much shorter than the parallel dimensions.  
This allows local Ohmic heating at the center of a current sheet to heat its
surroundings.  
With this additional physics, we have found it expedient to add a grid-scaled sixth-order 
hyperdiffusion to the momentum and temperature fields.
Hyperdiffusions are scaled locally to the grid spacing.
A logarithmic grid is specified by mapping a linear grid $g$ which ranges from $0$ to
$L_g$ to the physical coordinates $x$ by $x(g)=f_gL_g(\exp(g/(f_gL_g))-1)$ where $f_g$ is a factor which controls the grid stretching. 
The diffusion operator for a variable $f$  is then specified directly as 
$\partial f/\partial t = D_g \Delta g^4 \partial^6 f/\partial g^6$, where $\Delta g$ is the grid spacing increment in $g$, 
so that the diffusion scales with the grid.
We also artificially smooth the variation
in opacity through the addition of a finite relaxation time
physically motivated by the finite vaporization time of the dust grains.
Rosseland mean opacity is evolved by
$\partial \kappa /\partial t =  (1/t_\kappa)(\kappa_r -\kappa)$
where $\kappa$ is the opacity used in radiative transfer, $\kappa_r$ is the  
value of opacity derived from the opacity tables for the gas-dust mixture,
and $t_\kappa$ is a time constant.
These alterations maintain the integrability of the system of equations, 
and we have checked that the numerical smoothing coefficients are 
small enough to not significantly 
affect the results presented here. Our initial configuration is the same as in Paper~I.

The Rosseland mean opacities $\kappa_r$ that we use are an updated version of the
\citet{2001ApJ...553..321D} model.
Importantly, they incorporate a distribution of dust particles
comprised of ice, organic, and silicate components.
The sublimation temperatures of the different components 
are taken to be independent of the grain size, so
that the opacity has a sharp transition downwards when these temperatures are reached.
This model includes: graphite and ``astronomical'' silicates, 
with optical constants and fractional abundances from 
\citet{2001ApJ...548..296W}, and water ice with optical properties
from \citet{1984ApOpt..23.1206W}. 
The grain size distribution is $n(r)\propto r^{-3.5}$ between 
$r_{min}=0.005~\mathrm{\mu m}$ and $r_{max}= 1~\mathrm{mm}$, appropriate for chondrule forming regions. 
The opacity is calculated using  Mie theory and assuming grains are spheres. 
The sublimation temperatures for silicates and 
water ice grains are taken from \citet{1994ApJ...421..615P}, 
and for graphite, we assume $T_{sub}=1200~\mathrm{K}$. 
For the gas component, which is important when dust sublimates, we assume LTE.
The dust-to-gas mass ratio for silicates is $4\times10^{-3}$, for
organics, $2.5\times10^{-3}$, and for water ice, $4.7\times10^{-3}$, so
in total it is $1.21\times10^{-2}$.
Our model assumes a constant mean molecular mass $\mu=2.32$, 
which is not strictly consistent with the vaporization of solids.  
However,
we stop calculations before the dissociation of H$_2$ introduces a significant effect. 

As discussed in Paper~I, the short-circuit instability requires a strong
negative dependence of the resistivity on temperature.  
We start our simulations with $T=T_0=850~\mathrm{K}$,
where the ionization already has a rapid dependence on temperature even outside the current sheet.

In Figure~\ref{fig_class2r3} we show the evolution of 
 a current sheet model
with an initial ratio of thermal to magnetic pressure
at the box edge of  $\beta=1.5$,
background temperature $T_0=850~\mathrm{K}$, and background density
$\rho_0=10^{-7}$~g~cm$^{-3} $ with $B\propto \tanh(x/5\times
10^{10}\mathrm{\ cm} )$.
In this configuration, the shortest wavelength for MRI, based on the length 
scale where the magnetic Reynolds number is unity, is on the order of $1.7\times10^{11}\ \mathrm{cm}$.
The grid had $1000$ points, $L_g=5\times10^8~\mathrm{cm}$ and $f_g=0.09$.
Values of  $D_g=10^9~\mathrm{cm^2~s^{-1}}$ in the temperature equation, 
and $D_g = 10^{10}~\mathrm{cm^{2}~s^{-1}} $ in the momentum equation were used.
The opacity was smoothed with  $t_\kappa = 2\times10^3~\mathrm{s}$.
 Notice in particular
the small width of the narrowed current sheet.
As the opacity drops, the inverse mean free path of photons $\chi$ (plotted in
units of $\mathrm{cm^{-1}}$ in Figure~\ref{fig_class2r3}) decreases dramatically, and an
optically thin bubble
 is formed around the
current sheet as the temperature where silicates become liquid is passed.
The low opacity bubble has small internal temperature gradients, and hence smaller resistivity gradients.
As the resistivity gradient weakens, the instability shuts off. 
In this way the instability self-regulates.
For the parameters in Figure~\ref{fig_class2r3}, it
produces heated regions with temperatures that
only slightly exceed the melting temperature of the silicate grains
that dominate the opacity, 
the largest of which produce chondrules; at this point, the instability is saturated by the increased radiative cooling.
This is the most obvious configuration leading to chondrule formation.

If the current sheet is initially thinner, the heating rate becomes fast 
enough  that the heating is not affected
by the drop in opacity.
Figure~\ref{fig_class2r4} shows 
the same configuration as before,
except with a factor of five thinner initial current sheet 
$B\propto \tanh(x / 10^{10}\mathrm{\ cm} )$.
The grid had $800$ points, $L_g=10^8~\mathrm{cm}$ and $f_g=0.09$.
Values of  $D_g=10^8~\mathrm{cm^2~s^{-1}}$ in the temperature equation, 
and $D_g = 10^{10}~\mathrm{cm^2~s^{-1}} $ in the momentum equation were used.
The opacity was smoothed with $t_\kappa = 1\times10^2~\mathrm{s}$.
In this case, the short circuit instability narrows the current sheet until the 
 potassium ionization fraction (${x_e/a}$) reaches unity and the resistivity gradient disappears. 
The current sheet continues to heat as it dissipates, although we stop our calculation as it 
 breaks down because we do not include the effects of $H_2$ dissociation 
 and other changes to the effective equation of state above $2000\mathrm{\ K}$.

\section{Discussion and Conclusions} \label{sec_discussion}

The models in this paper are  far from a full treatment.  
However,  they demonstrate that 
the short-circuit instability is a plausible candidate for
chondrule formation as 
 shown by the model in Figure~\ref{fig_class2r3}, which starts within the
range of boundary conditions for current sheets seen in our global models of
protoplanetary disks, and climbs to chondrule melting temperatures within
hundreds of hours.
For thinner initial current sheets, the instability can 
reach temperatures high enough to fully ionize potassium,
which is also high enough to produce barred olivine textures.
This behavior is also compatible with the remelting of Type B CAIs.
As the formation of the first chondrules and CAIs overlaps in 
time \citep{2007ApJ...671L.181M,2009LPI....40.2006Y,Connelly2012},
it is plausible that the 
same mechanism is responsible for processing Type B CAIs and the oldest chondrules.

We have also noted that the drop in opacity associated with the destruction 
of organic components, at $1200~\mathrm{K}$ in our opacity model, can produce a 
behavior similar to that in Figure~\ref{fig_class2r3}, but at temperatures below the melting point of silicates.
The temperatures produced in this type of event would be appropriate for quickly 
annealing amorphous silicate dust to produce crystalline silicates.

Chondrule formation by the melting of precursor grains in reconnection regions subject to the 
short-circuit instability of Paper~I seems compatible with 
constraints from chondrule-matrix complementarity, chondrule-chondrule complementarity, and
multiple heating listed in Section~\ref{sec_intro}, as it is a local process that can occur
many times to the same body of meteorite parent body precursor material.

As the MRI generates magnetic fields near equipartition with the gas thermal
energy, compressive heating is a substantial 
contributor to the temperature.  Our
initial condition, as in Paper~I, is chosen to be adiabatic-hydrostatic, which
is the gentlest configuration we have tested.
Determining the
strength of the compressive heating, and its ability to meet the threshold
temperature for the short-circuit instability, 
 will require models that
self-consistently form the current sheets in a large scale
simulation that includes the action of temperature-dependent resistivity. 
Such simulations will also allow determination of
the correct physical length scale for the current sheet thickness
at the onset of the instability.
This length scale will strongly 
affect the heating and cooling times.

Furthermore, multidimensional studies of the current sheets are needed, to
establish how they might break up (due to plasmoid instabilities or buckling of
the current sheet) and hence the size and lifetime 
of the heated regions, which will be
important for determining the cooling rates of chondrules in this scenario. 
Finally, we note that the temperatures explored here have been discussed
in reference only to chondrule and CAI formation.  
However, given 
more complete ionization models, other high temperature
mineral processing, such as the crystallization of silicates, destruction of organics, 
or sublimation of ices, may occur through the same mechanism.

\acknowledgements 
We thank Remo Collet for useful discussions.
A.H. was partly
supported by a Kalbfleisch Fellowship 
from the American Museum of Natural History. We also acknowledge support
from the NSF through CDI grant AST08-35734, AAG grant AST10-09802, and NASA through COS grant NNX10AI42G (DSE).
PD acknowledges a grant from PAPIIT-DGAPA.

\bibliographystyle{apj}

\end{document}